\documentclass[12pt]{iopart}
\usepackage{units}
\usepackage{graphicx}
\usepackage{cite}
\usepackage{amssymb}

\usepackage{textcomp}

\begin{document}
\title[Magnetoelectric properties of $A_2$\lbrack FeCl$_5$(H$_2$O)\rbrack]{Magnetoelectric properties of $A_2$[FeCl$_5$(H$_2$O)] with $A=$~K, Rb, Cs}

\author{M Ackermann$^{1}$, T Lorenz$^{2}$,
P Becker$^{1}$ and L Bohat\'y$^1$}

\address{$^1$ Institut f\"ur Kristallographie, Universit\"at zu K\"oln, Greinstra\ss e 6, 50937 K\"oln, Germany}
\address{$^2$ II. Physikalisches Institut, Universit\"at zu K\"oln, Z\"ulpicher Stra\ss e 77, 50937 K\"oln, Germany}

\ead{ladislav.bohaty@uni-koeln.de}

\begin{abstract}
The compounds $A_2$[FeCl$_5$(H$_2$O)] with $A=$~K, Rb, Cs are identified as new linear magnetoelectric (non-multiferroic) materials. We present a detailed investigation of their linear magnetoelectric properties by measurements of  pyroelectric currents, dielectric constants and magnetization.  The anisotropy of the linear magnetoelectric effect of the K-based and Rb-based compound is consistent with the magnetic point group $m'm'm'$, already reported in literature. A symmetry analysis of the magnetoelectric effect of the Cs-based compound allows to determine the magnetic point group $mmm'$ and to develop a model for its magnetic structure. 
In addition, magnetic-field versus temperature phase diagrams are derived and compared to the closely related multiferroic (NH$_4$)$_2$[FeCl$_5$(H$_2$O)].
\end{abstract}

\submitto{{\rm This is an author-created, un-copyedited version of an article published in: \\{\it J.~Phys.: Condens. Matter} {\bf26} 506002 (2014) \\ IOP Publishing Ltd is not responsible for any errors or omissions in this version of the manuscript or any version derived from it. The Version of Record is available online at: \\ http://dx.doi.org/10.1088/0953-8984/26/50/506002}}
\maketitle

\section{Introduction}
\label{intro}

Recently, (NH$_4$)$_2$[FeCl$_5$(H$_2$O)] was established as new multiferroic material with a strong magnetoelectric coupling and with rather complex magnetic-field versus temperature phase diagrams~\cite{Ackermann2013}. It belongs to the large family of erythrosiderite-type compounds $A_2$[Fe$X_5$(H$_2$O)], where $A$ stands for an alkali metal or ammonium ion and $X$ for a halide ion. The large variety of structurally closely related compounds within this family opens the possibility to study the structural prerequisites for the occurrence of multiferroicity in this class of materials and the impact of crystal-chemical modifications on multiferroic and magnetoelectric properties in general. Here, we present a detailed study of the magnetoelectric properties of the alkali-based compounds K$_2$[FeCl$_5$(H$_2$O)], Rb$_2$[FeCl$_5$(H$_2$O)] and Cs$_2$[FeCl$_5$(H$_2$O)].

\begin{table}[b]
\centering
\begin{tabular}{ccccc}
\hline
crystal: & (NH$_4$)$_2$[FeCl$_5$(H$_2$O)] & K$_2$[FeCl$_5$(H$_2$O)] &  Rb$_2$[FeCl$_5$(H$_2$O)] &  Cs$_2$[FeCl$_5$(H$_2$O)] \\ 
\hline
$a$ (\r{A}): & 13.706 & 13.75 & 13.825 & 7.426 \\

$b$ (\r{A}): & 9.924 & 9.92 & 9.918 & 17.306 \\

$c$ (\r{A}): & 7.024 & 6.93 & 7.100 & 8.064 \\

\hline
$T_{\rm N}$ (K): & 7.25 & 14.06 & 10.05 & 6.5 \\
\hline
\end{tabular}
\label{tab:structur}
\caption{Room-temperature lattice constants and N\'eel temperatures of $A_2$[FeCl$_5$(H$_2$O)] with $A=$~(NH$_4$), K, Rb, Cs, taken from~\cite{McElearny1978, Figgis1978, Lindquist1947, Connor1979, Greedan1980}.}
\end{table}

\begin{figure}[t]
\includegraphics[width=\textwidth]{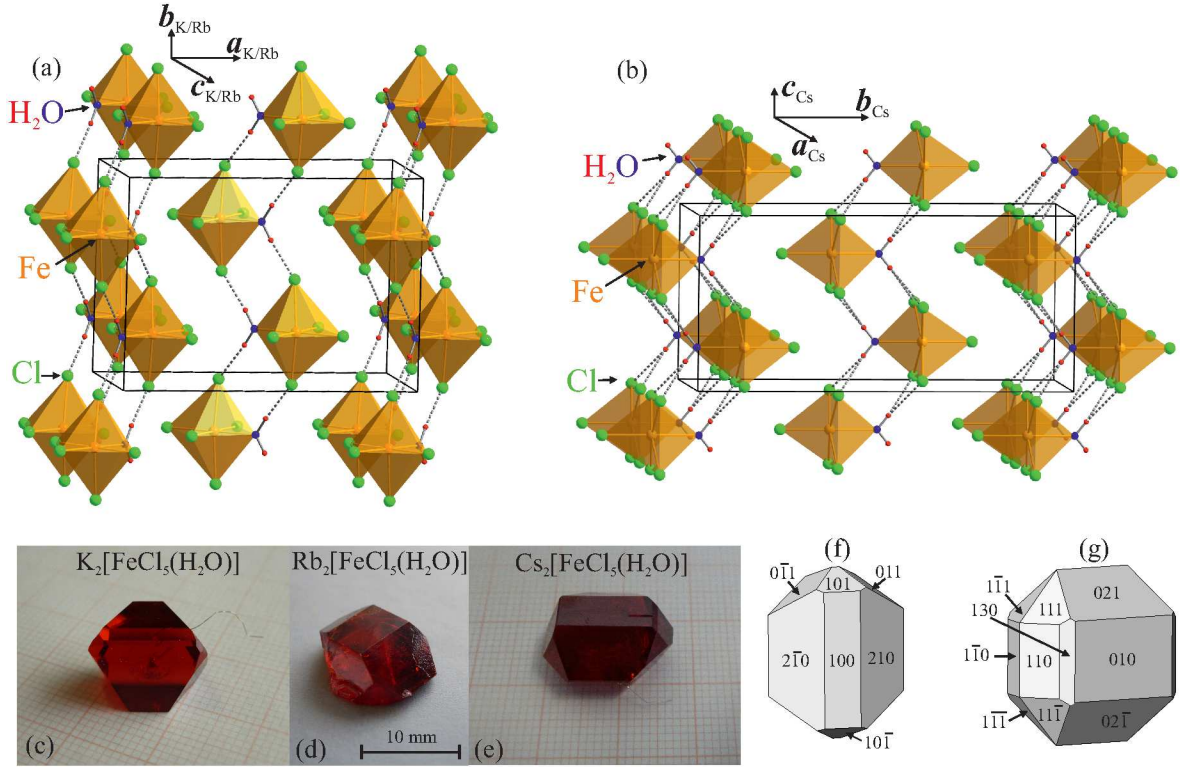}
\caption{$A_2$[FeCl$_5$(H$_2$O)] crystal structures for (a) $A=$~(NH$_4$), K, Rb and (b) $A=$~Cs. Note that the positions of the $A^+$ ions are omitted for clarity. The structural data are taken from~\cite{Lindquist1947, Greedan1980}. (c)-(e) Grown $A_2$[FeCl$_5$(H$_2$O)] crystals for $A=$~K, Rb and Cs . Typical morphologies of $A_2$[FeCl$_5$(H$_2$O)] crystals with (f) $A=$~(NH$_4$), K, Rb and (g) $A=$~Cs. }
\label{figure1}
\end{figure}

The erythrosiderite-type compounds form a series of antiferromagnets with N\'eel temperatures ranging from  \unit[6]{K} to \unit[15]{K}~\cite{Luzon2008}. The room-temperature crystal structures of $A_2$[FeCl$_5$(H$_2$O)] are orthorhombic with the space group $\textit{Pnma}$ for $A=$~(NH$_4$), K, Rb and $\textit{Cmcm}$ for $A=$~Cs, respectively~\cite{Lindquist1947, Bellanca1948, Figgis1978, Connor1979, Greedan1980, Schulz1995, Lackova2013}. Both structure types are closely related but not isomorphic, see figures~\ref{figure1}\,(a) and (b). In both cases the structure consists of isolated $A^+$ units and isolated complex groups [FeCl$_5$(H$_2$O)]$^{2-}$ of sixfold octahedrally coordinated iron(III). The unit cells contain eight symmetrically equivalent $A^+$ cations and four [FeCl$_5$(H$_2$O)]$^{2-}$ octahedra.  Besides ionic bonds between the structural building blocks, there are H-bonds \mbox{(via O--H--Cl)} between neighbouring [FeCl$_5$(H$_2$O)]$^{2-}$ octahedra, which further stabilizes the crystal structures. These H-bonded octahedra form zigzag chains, which run along $\bi{b}$ for $A=$~(NH$_4$), K, Rb or along $\bi{c}$ for $A=$~Cs.  Along these chains the Fe--O bonds of adjacent octahedra are oriented mutually antiparallel to each other~\cite{Connor1979, Greedan1980, McElearny1978}. For $A=$~(NH$_4$), K, Rb, the Fe--O bonds of the octahedra are  approximately lying parallel to the $ac$ plane with alternating angles in the order of $\pm 40^{\circ}$ relative to the $\bi{a}$ axis. For $A=$~Cs, these bonds are oriented parallel to $\bi{b}$.  Taking the zigzag chains as the dominant feature of the two crystal-structure types, the axes $\{\bi{a}, \bi{b}, \bi{c}\}$ in the $Pnma$ structure corresponds to the axes $\{\bi{b}, \bi{c}, \bi{a}\}$ in the $Cmcm$ structure, see figures~\ref{figure1}\,(a) and (b). The lattice constants are summarized in table~\ref{tab:structur}.

The magnetic ordering phenomena in the $A_2$[FeCl$_5$(H$_2$O)] series have been subject of various investigations in the past. In this context, N\'{e}el temperatures ranging from  \unit[6]{K} to \unit[15]{K} have been derived from measurements of the magnetic susceptibility~\cite{McElearny1978, Connor1979}, see table~\ref{tab:structur}. There are some clear differences between (NH$_4$)$_2$[FeCl$_5$(H$_2$O)] and the corresponding alkali-based compounds with $A= {\rm K}$, Rb, Cs. Susceptibility data of the alkali-based compounds identify the $\bi{a}$ axis as the magnetic easy axis, while no easy axis can be derived from the susceptibility data of (NH$_4$)$_2$[FeCl$_5$(H$_2$O)]~\cite{McElearny1978, Connor1979}. Moreover, heat-capacity measurements reveal a second phase transition at $T_{\rm C}\simeq\unit[6.87]{K}$ for (NH$_4$)$_2$[FeCl$_5$(H$_2$O)]~\cite{McElearny1978}, which coincides with the onset of a spontaneous electric polarization~\cite{Ackermann2013}. In contrast, for the alkali-based compounds only single transitions are reported~\cite{McElearny1978,Puertolas1985}. These observations, especially the discovery of multiferroicity, imply the presence of a complex magnetic spin structure in (NH$_4$)$_2$[FeCl$_5$(H$_2$O)] below $\unit[6.87]{K}$, but a determination of the magnetic structure is still missing. In contrast, the magnetic structures of the alkali-based compounds with $A= {\rm K}$, Rb were determined via neutron scattering~\cite{Gabas1995a, Luzon2008a}. The antiferromagnetically ordered phases of both compounds are described by the magnetic space group $Pn'm'a'$. The spin direction alternates between $\pm\bi{a}$ along the zigzag chains of the [FeCl$_5$(H$_2$O)]$^{2-}$ octahedra running along $\bi{b}$ and the spins of neighbouring chains are in phase.

The paper is organized as follows. First, the crystal growth of $A_2$[FeCl$_5$(H$_2$O)] with $A=$~K, Rb, Cs and the experimental techniques for the study of their magnetoelectric properties are described. Then, the results of the magnetic-susceptibility measurements and the dielectric investigations are presented and discussed in detail. The magnetoelectric effect and its temperature dependence of all compounds is analysed. A detailed symmetry analysis of the magnetoelectric effect of the Cs-based compound allows to develop a model for its magnetic structure with the magnetic point group $mmm'$. The paper is concluded by a discussion of the magnetic-field versus temperature phase diagrams of all compounds. Finally, the magnetoelectric properties of the alkali-based compounds are compared to those of the related multiferroic (NH$_4$)$_2$[FeCl$_5$(H$_2$O)].

\section{Experiments}

\begin{table}[t]
\centering
\begin{tabular}{ccc}
\hline
\bf{crystal} & \bf{growth temperature} & \bf{starting composition of growth solution} \\ 
\hline

K$_2$[FeCl$_5$(H$_2$O)] & \unit[311]{K} & 1KCl~+~2.5FeCl$_3$ (non-stoichiometric ratio) \\

Rb$_2$[FeCl$_5$(H$_2$O)] & \unit[323]{K} & 2RbCl~+~1FeCl$_3$ (stoichiometric ratio) \\

Cs$_2$[FeCl$_5$(H$_2$O)] & \unit[323]{K} & 2CsCl~+~1FeCl$_3$ (stoichiometric ratio) \\
\hline
\end{tabular}
\label{tab:growth}
\caption{Parameters for crystal growth from aqueous solution of $A$Cl and FeCl$_3$ with a surplus of HCl by controlled evaporation of the solvent; growth period: 8-12 weeks.}
\end{table}

Growth of large single crystals $A_2$[FeCl$_5$(H$_2$O)] ($A=$ K, Rb, Cs) was achieved by solution growth using aqueous solutions of $A$Cl and FeCl$_3$ with a surplus of HCl. During the crystal-growth process the evaporation of the solvent was controlled. The growth temperatures as well as the starting compositions of the growth solutions for the different compounds are summarized in table~\ref{tab:growth}. Growth periods of typically 6--8 weeks yielded optically clear, red single crystals with well-developed flat morphological faces and with maximal dimensions of \mbox{$\simeq40\times$30$\times\unit[20]{mm^{3}}$}, see figure~\ref{figure1}\,(c)-(e). The typical morphology of the crystals is displayed in figure~\ref{figure1}\,(f) and (g). Oriented $(100)$, $(010)$ and $(001)$ samples were prepared using the morphological faces as reference planes. For the magnetic-susceptibility measurements, samples with typically a thickness of $\sim$\,$\unit[1]{mm}$ and surfaces of $\sim$\,$\unit[5]{mm^2}$ were used. For the dielectric investigations, plate-like samples with typical surfaces of ($\sim$\,$\unit[30]{mm^2}$) were used, which were vapour-metallized with silver electrodes.

The investigations of magnetization were performed in a temperature range of $2-\unit[300]{K}$ and a magnetic-field range of $0-\unit[7]{T}$  with a commercial SQUID magnetometer (MPMS, Quantum Design). The dielectric properties were investigated in the same temperature range in a cryostat equipped with a \unit[15]{T} magnet. The relative dielectric constants $\epsilon_{i}^{r}$ ($i = a, b, c$) were determined from the capacitances, recorded by a capacitance bridge (Andeen-Hagerling 2500A) at a frequency of \unit[1]{kHz}. A time-integration of the pyroelectric currents, recorded by an electrometer (Keithley 6517) led to the temperature dependence of the magnetoelectrically induced electric polarization. Static electric poling fields of at least \unit[200]{V/mm}, which were applied well above the transition temperature, enforced single magnetoelectric domain states during cooling the crystal to base temperature. The pyroelectric current was measured during the following  heating process with removed poling field and with a constant heating rate of \unit[3]{K/min}. The magnetoelectric domains and hence the electric polarization could be completely switched in most of the cases by reversing the electric poling field. The magnetic-field dependence of the electric polarization was obtained by resorting the data of the temperature-dependent measurements.

\section{Results and discussion}

\subsection*{Magnetic properties}
\label{sec:magpropa2}

\begin{figure}[t]
\includegraphics[width=\textwidth]{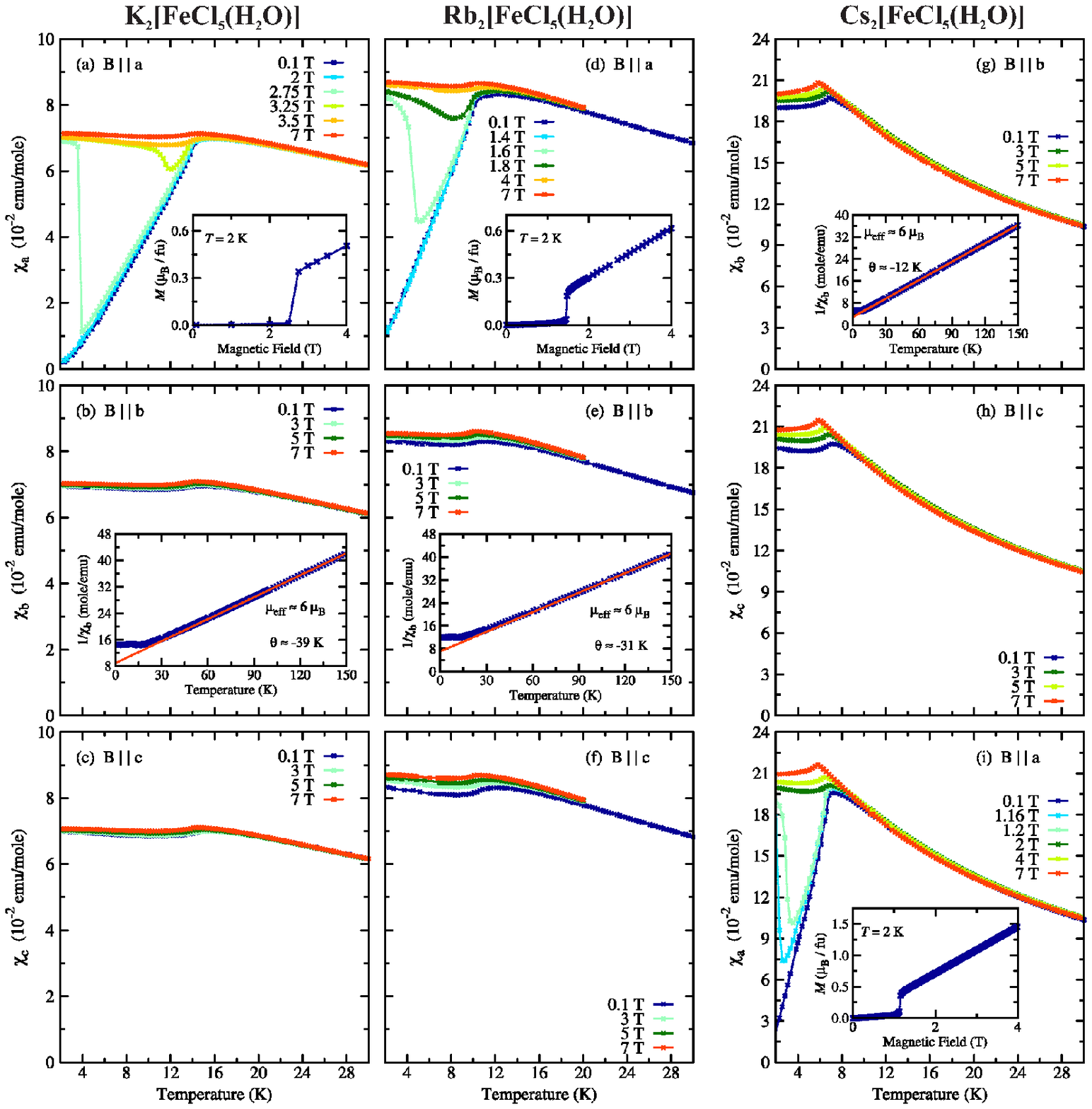}
\caption{Temperature dependences of the magnetic susceptibility $\chi_{a}$, $\chi_{b}$, and $\chi_{c}$ of $A_2$[FeCl$_5$(H$_2$O)] for $A=$~K (a)--(c), $A=$~Rb (d)--(f), $A=$~Cs (g)--(i) for different magnetic fields parallel to $\bi{a}$, $\bi{b}$ and $\bi{c}$. The insets in (a), (d) and (i) show the magnetization at \unit[2]{K} for a magnetic field parallel to $\bi{a}$. The insets in (b), (e) and (g) show the temperature-dependent inverse susceptibilty $1/\chi_{\mathrm{b}}$ together with Curie-Weiss fits (lines) in the temperature range from \unit[100]{K} to \unit[300]{K}. Note that the $\{\bi{a}, \bi{b}, \bi{c}\}$ axes in the $Pnma$ structure for $A=$~K, Rb  correspond to the $\{\bi{b}, \bi{c}, \bi{a}\}$ axes in the $Cmcm$ structure for $A=$~Cs, see figure~\ref{figure1}.}
\label{figure2}
\end{figure}

The magnetic-susceptibility measurements of $A_2\lbrack$FeCl$_5$(H$_2$O)$\rbrack$ with $A=$~K, Rb, Cs are summarized in figure~\ref{figure2}. The magnetic properties of all three compounds are rather similar. The low-field curves $\chi(T)$ show well-defined kinks at $T_{\rm N}^{\mathrm{K}}\simeq \unit[14.3]{K}$, $T_{\rm N}^{\mathrm{Rb}}\simeq \unit[10.2]{K}$ and $T_{\rm N}^{\mathrm{Cs}}\simeq \unit[6.8]{K}$ for $A=$~K, Rb and Cs, respectively, in agreement with previous results~\cite{ Connor1979, McElearny1978, Gabas1995a, Puortelas1982, Carlin1977, Filho1978}. In all three cases, $\chi_a$ decreases below $T_{\rm N}$  with decreasing temperature  and approaches zero for $T\rightarrow\unit[0]{K}$, while $\chi_b$ and $\chi_c$ hardly change, indicating that the spins are oriented parallel to the $\bi{a}$ axis, the magnetic easy axis. Note, however, that with respect to the zigzag chains of [FeCl$_5$(H$_2$O)]$^{2-}$ octahedra the $\bi{a}$ axis is oriented differently in the two structure types, see figure~\ref{figure1}. Therefore, the magnetic easy axis in the Cs-based compound is perpendicular to that in the K-based and Rb-based compound.  In the K-based and Rb-based compound $T_{\rm N}^{\mathrm{K}}$ and $T_{\rm N}^{\mathrm{Rb}}$ slightly decrease with increasing magnetic field independently from the field direction. In contrast, in the Cs-based compound $T_{\rm N}^{\mathrm{Cs}}$ strongly decreases for all three field directions with increasing magnetic fields, which is in agreement with results of a M\"ossbauer study~\cite{Johnson1987}.

For larger magnetic fields parallel to $\bi{a}$, the decrease of $\chi_a(T)$ below $T_{\rm N}$  systematically vanishes for all compounds. Above the critical fields $B_{\mathrm{SF}}^{\mathrm{K}}\simeq \unit[3.5]{T}$, $B_{\mathrm{SF}}^{\mathrm{Rb}}\simeq \unit[1.5]{T}$ and $B_{\mathrm{SF}}^{\mathrm{Cs}}\simeq \unit[1.2]{T}$ the characteristics of $\chi_i(T)$ become nearly identical in all the compounds for all three magnetic-field directions $\bi{a}$, $\bi{b}$,~$\bi{c}$, see figure~\ref{figure2}\,(a),~(d),~(i). Therefore, spin-flop transitions occur at $B_{\mathrm{SF}}^{A}$, with a rotation of the spins from being oriented along the $\bi{a}$ axis to lying within the plane that is perpendicular to $\bi{a}$.	These spin-flop transitions can also be seen in the low-temperature measurements of the magnetization as a function of the magnetic field, see  the insets in figure~\ref{figure2}\,(a),~(d),~(i). Note that the field-dependent magnetization of the K-based compound was obtained by resorting the temperature-dependent magnetization data.

The inverse susceptibilities follow a Curie-Weiss behaviour from about \unit[50]{K} up to room temperature for all three compounds, see the insets in  figure~\ref{figure2}\,(b),~(e),~(g). Here, only the data of $1/\chi_b$ are displayed, because the data of $1/\chi_a$ and $1/\chi_c$ for all the compounds are almost identical in the high-temperature regime. Linear fits to the data of $1/\chi_i^{A}$ for $T>\unit[100]{K}$ yield negative Weiss temperatures $\theta^{A}\simeq -\unit[39]{K},  -\unit[31]{K}, -\unit[12]{K}$ for $A=$~K, Rb and Cs, respectively, and an effective magnetic moment $\mu_{\mathrm{eff}}\simeq \unit[6]{\mu_{\mathrm{B}}}$. The negative Weiss temperatures signal net antiferromagnetic exchange interactions, whose magnitude roughly scales with $T_{\rm N}$. With increasing ionic radius of the alkali ions the sum of the exchange interactions diminishes, most probably due to the increasing distance of the magnetic ions. The value of the effective magnetic moment lies close to $\mu_{\mathrm{eff}}=g\mu_{B}\sqrt{S(S+1)}=\unit[5.92]{\mu_{B}}$, as expected for a 3d$^5$ high-spin configuration of the Fe$^{3+}$ ions and $g=2$.

\subsection*{Dielectric properties}

\begin{figure}[t]
\includegraphics[width=\textwidth]{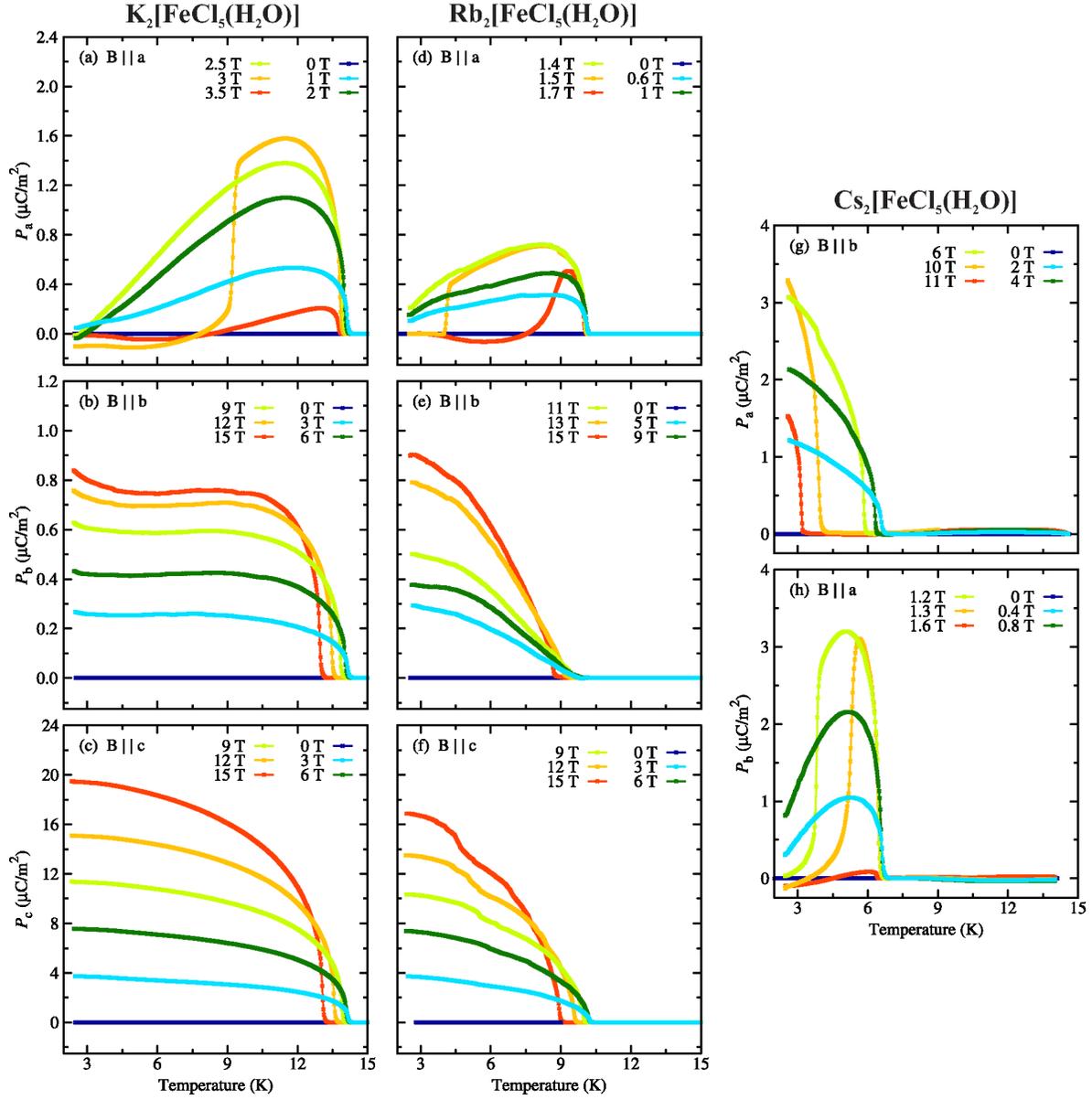}
\caption{Temperature dependences of the electric polarization $P_a$, $P_b$ and $P_c$ of \mbox{(a)--(c) K$_2$[FeCl$_5$(H$_2$O)],} (d)--(f) Rb$_2$[FeCl$_5$(H$_2$O)] and (g)--(h) Cs$_2$[FeCl$_5$(H$_2$O)] for magnetic fields applied parallel to the $\bi{a}$, $\bi{b}$ or $\bi{c}$ axis, respectively. The magnetic-field configurations, where no electric polarization is induced, are not presented here.}
\label{figure4}
\end{figure}

\begin{figure}[t]
\includegraphics[width=\textwidth]{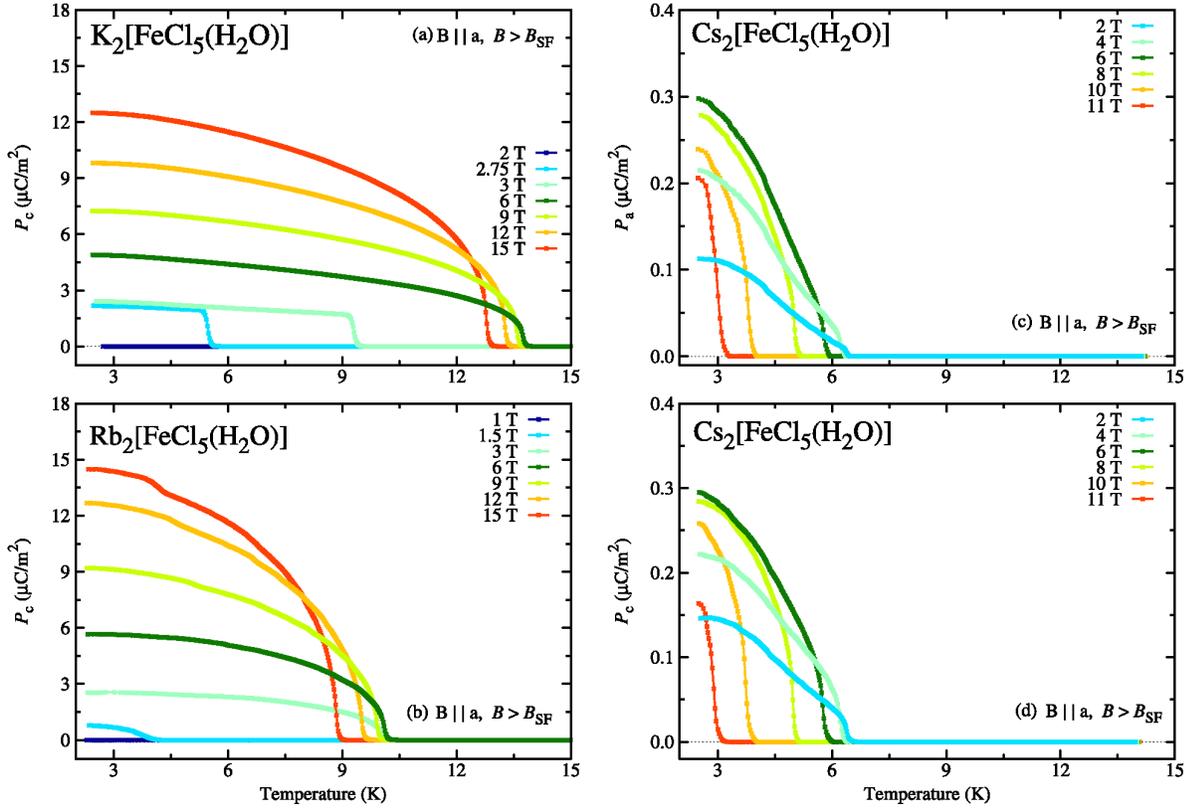}
\caption{Temperature dependences of the electric polarization $P_a$ and $P_c$ of (a) K$_2$[FeCl$_5$(H$_2$O)], \mbox{(b) Rb$_2$[FeCl$_5$(H$_2$O)]} and (c)--(d) Cs$_2$[FeCl$_5$(H$_2$O)] for magnetic fields applied parallel to the $\bi{a}$ axis above the spin-flop transitions.}
\label{figure5}
\end{figure}

Figure~\ref{figure4} displays the temperature-dependent measurements of the electric polarization $\bi{P}=(P_a,P_b,P_c)$ of $A_2$[FeCl$_5$(H$_2$O)] with $A=$~K, Rb and Cs in magnetic fields applied parallel to $\bi{a}$, $\bi{b}$ and $\bi{c}$. In none of the three compounds, an electric polarization arises  below the corresponding N\'eel temperature in zero magnetic field. Applied  magnetic fields, however, induce an electric polarization in all cases. Therefore, the three compounds are not ferroelectric (multiferroic) but show a magnetoelectric effect. The magnetic-field configurations, where no electric polarization is induced, are not presented here. The induced components of the electric polarization of K$_2$[FeCl$_5$(H$_2$O)] and Rb$_2$[FeCl$_5$(H$_2$O)] behave very similar with respect to their temperature and magnetic-field dependences. The induced electric polarization of Cs$_2$[FeCl$_5$(H$_2$O)] has slightly different characteristics.

As can be seen in figure~\ref{figure4}\,(a) and (d), the polarization components $P_a^{\mathrm{K}}$ and $P_a^{\mathrm{Rb}}$ in the K-based and Rb-based compound reach broad maxima around \unit[12]{K} and \unit[9]{K}, respectively, for $\bi{B}||\bi{a}$ and then approach zero for $T\rightarrow\unit[0]{K}$. With increasing field strength $P_a^{\mathrm{K}}$ and $P_a^{\mathrm{Rb}}$ grow linearly in the low-field range until they are suppressed completely in the vicinity of the spin-flop transitions at \mbox{$B_{\mathrm{SF}}^{\mathrm{K}}=\unit[3.5]{T}$} and $B_{\mathrm{SF}}^{\mathrm{Rb}}=\unit[1.5]{T}$, respectively. At higher field strengths, electric polarizations $P_c$ parallel to $\bi{c}$ occur, which also grow linearly with further increasing magnetic fields, see figures~\ref{figure5}\,(a) and (b). For both compounds, the temperature dependences of $P_b$ and $P_c$ for $\bi{B}||\bi{b}$ and $\bi{B}||\bi{c}$, respectively, are very similar, but different from $P_a$ for $\bi{B}||\bi{a}$, see figure~\ref{figure4}\,(b), (c), (e) and (f). In both cases the induced polarizations depend again linearly on the field strength and $P_c$ is about one order of magnitude larger than $P_b$.

In Cs$_2$[FeCl$_5$(H$_2$O)], a magnetic field parallel to $\bi{a}$ induces an electric polarization along $\bi{b}$, while a magnetic field parallel to $\bi{b}$ causes an electric polarization along $\bi{a}$, see figure~\ref{figure4}\,(g),~(h). The temperature dependence of $P_b^{\mathrm{Cs}}$ for $\bi{B}||\bi{a}$ is very similar to that of $P_a^{\mathrm{K}}$ and $P_a^{\mathrm{Rb}}$ for $\bi{B}||\bi{a}$. It reaches a broad maximum around \unit[5]{K} and then approaches zero for $T\rightarrow\unit[0]{K}$. In the vicinity of the spin-flop transition at $B_{\mathrm{SF}}^{\mathrm{Cs}}=\unit[1.2]{T}$, $P_b^{\mathrm{Cs}}$ is suppressed completely and instead an electric polarization with the components $P_a^{\mathrm{Cs}}$ and $P_c^{\mathrm{Cs}}$ arises, see figure~\ref{figure5}\,(c) and (d). For $\bi{B}||\bi{b}$ an electric polarization along $\bi{a}$ is created, which saturates quickly with decreasing temperature, and for $\bi{B}||\bi{c}$ no electric polarization emerges at all.

\begin{figure}[t]
\includegraphics[width=\textwidth]{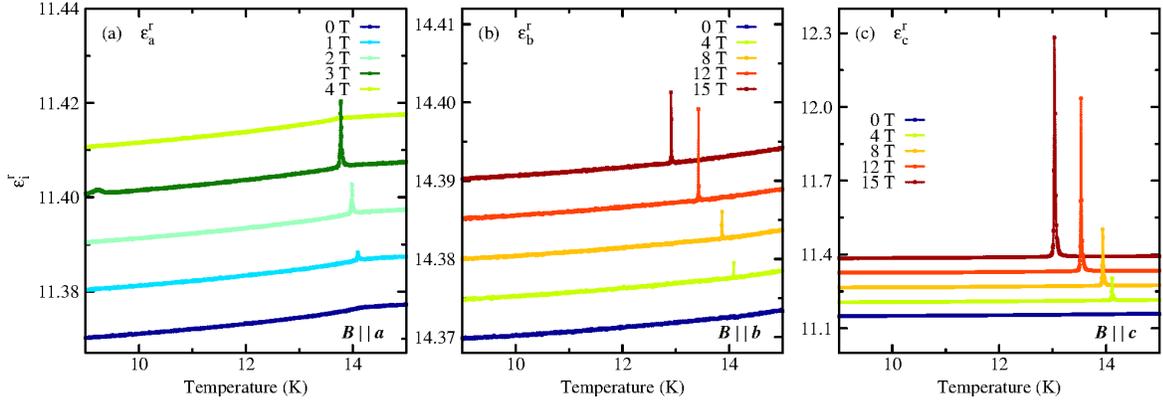}
\caption{Representative temperature dependences of the longitudinal components of the dielectric tensor $\epsilon^{r}_{i}$ ($i=a, b, c$ from left to right) of K$_2$[FeCl$_5$(H$_2$O)] for different magnetic fields applied parallel to $\bi{a}$, $\bi{b}$ or $\bi{c}$ (at a frequency of \unit[1]{kHz}). For clarity, with increasing field strength the curves are shifted with respect to each other by constant offsets of 0.01 in (a), 0.005 in (b) and 0.06  in (c).}
\label{figure6}
\end{figure}

The occurrence of the magnetic-field induced electric polarizations for $T<T_{\rm N}$ also causes anomalies in the temperature and magnetic-field dependences of the corresponding longitudinal components $\epsilon_i^r$ of the dielectric tensor. To illustrate this, representative measurements of the temperature-dependent $\epsilon_i^r$ ($i=a,b,c$) of K$_2$[FeCl$_5$(H$_2$O)] for $\bi{B}$ parallel to $\bi{a}$, $\bi{b}$ and $\bi{c}$ (at a frequency of \unit[1]{kHz}) are displayed in figure~\ref{figure6}. The zero-field curves show only faint kinks at $T_{\rm N}$, but in finite magnetic fields, spiky anomalies occur at the transition temperatures, which grow in intensity with increasing magnetic field in all cases. For  $\bi{B}||\bi{a}$, above the magnetic spin-flop field, the spiky anomaly disappears again, while for the other field directions the anomalies stay present up to the maximum field strength of $\unit[15]{T}$. In the vicinity of the spin-flop field, a second anomaly occurs, which coincides with the suppression of the electric polarization component $P_a$. 

\subsection*{Magnetoelectric effect}

\begin{figure}[t]
\includegraphics[width=\textwidth]{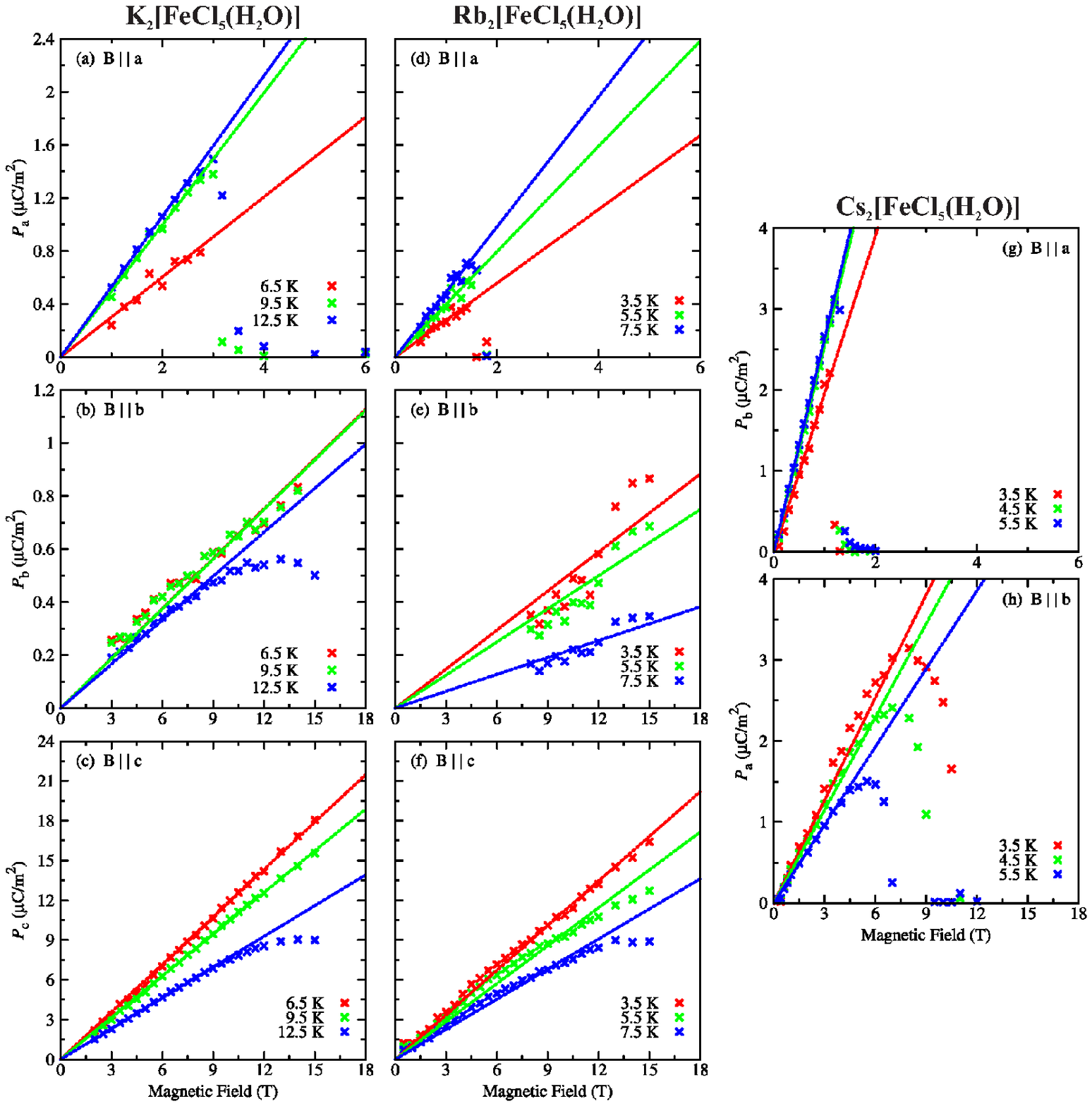}
\caption{Magnetic field dependences of the electric polarization of \mbox{(a)--(c) K$_2$[FeCl$_5$(H$_2$O)],} \mbox{(d)--(f) Rb$_2$[FeCl$_5$(H$_2$O)]} and (g)--(h) Cs$_2$[FeCl$_5$(H$_2$O)] at representative temperatures. The field-dependent data were obtained by resorting the temperature-dependent data of figure~\ref{figure4}. The solid lines are linear fits to the data for small magnetic-field strengths. Note that in the case of Cs$_2$[FeCl$_5$(H$_2$O)] no electric polarization is induced for $\bi{B}||\bi{c}$.}
\label{figure7}
\end{figure}

\begin{figure}[t]
\includegraphics[width=\textwidth]{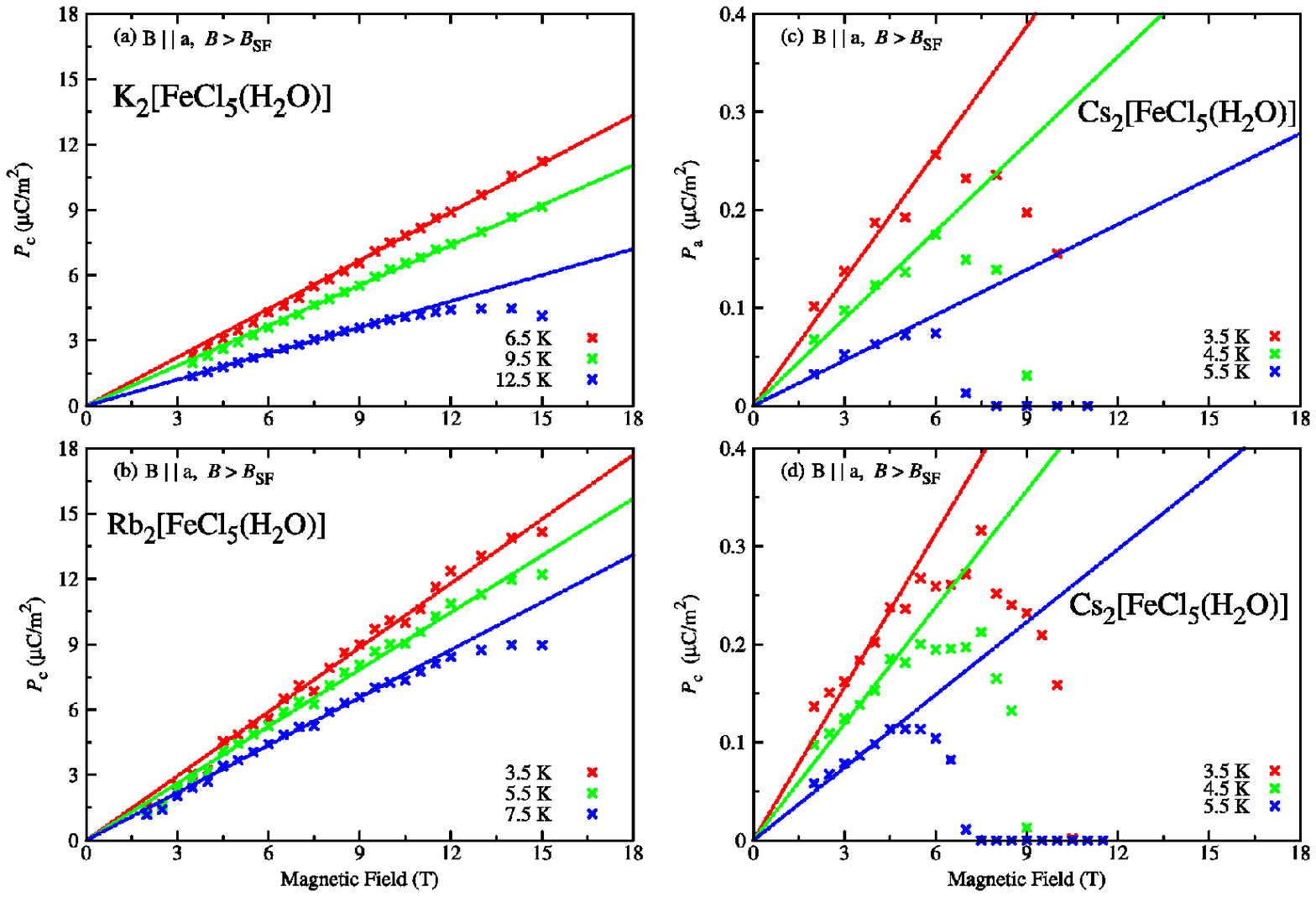}
\caption{Magnetic field dependences of the electric polarization of (a) K$_2$[FeCl$_5$(H$_2$O)], \mbox{(b) Rb$_2$[FeCl$_5$(H$_2$O)]} and (c)--(d) Cs$_2$[FeCl$_5$(H$_2$O)] at representative temperatures for applied magnetic fields along $\bi{a}$ above the spin-flop transition. The field-dependent data were obtained be resorting the temperature-dependent data of figure~\ref{figure5}. The solid lines are linear fits to the data for small magnetic field strengths.}
\label{figure8}
\end{figure}

For magnetic fields well below or above the spin-flop transitions, the magnetic-field induced electric polarization in $A_2$[FeCl$_5$(H$_2$O)] with $A=$~K, Rb, Cs depends in all cases linearly on the magnetic field strength. This indicates that all three compounds are linear magnetoelectrics.  The linear field dependence of the electric polarization of all compounds is illustrated in figures~\ref{figure7} and~\ref{figure8}, by plotting the components  $P_i^{\mathrm{K}}$, $P_i^{\mathrm{Rb}}$ and  $P_i^{\mathrm{Cs}}$ ($i=a,b,c$) as functions of the magnetic field at representative temperatures. Linear fits to the magnetic-field dependent electric polarization data yield the linear magnetoelectric tensor components \mbox{$\alpha_{ij}(T)=\partial \mu_0 P_i(T)/\partial B_j$}  in SI units.

\begin{figure}[t]
\includegraphics[width=\textwidth]{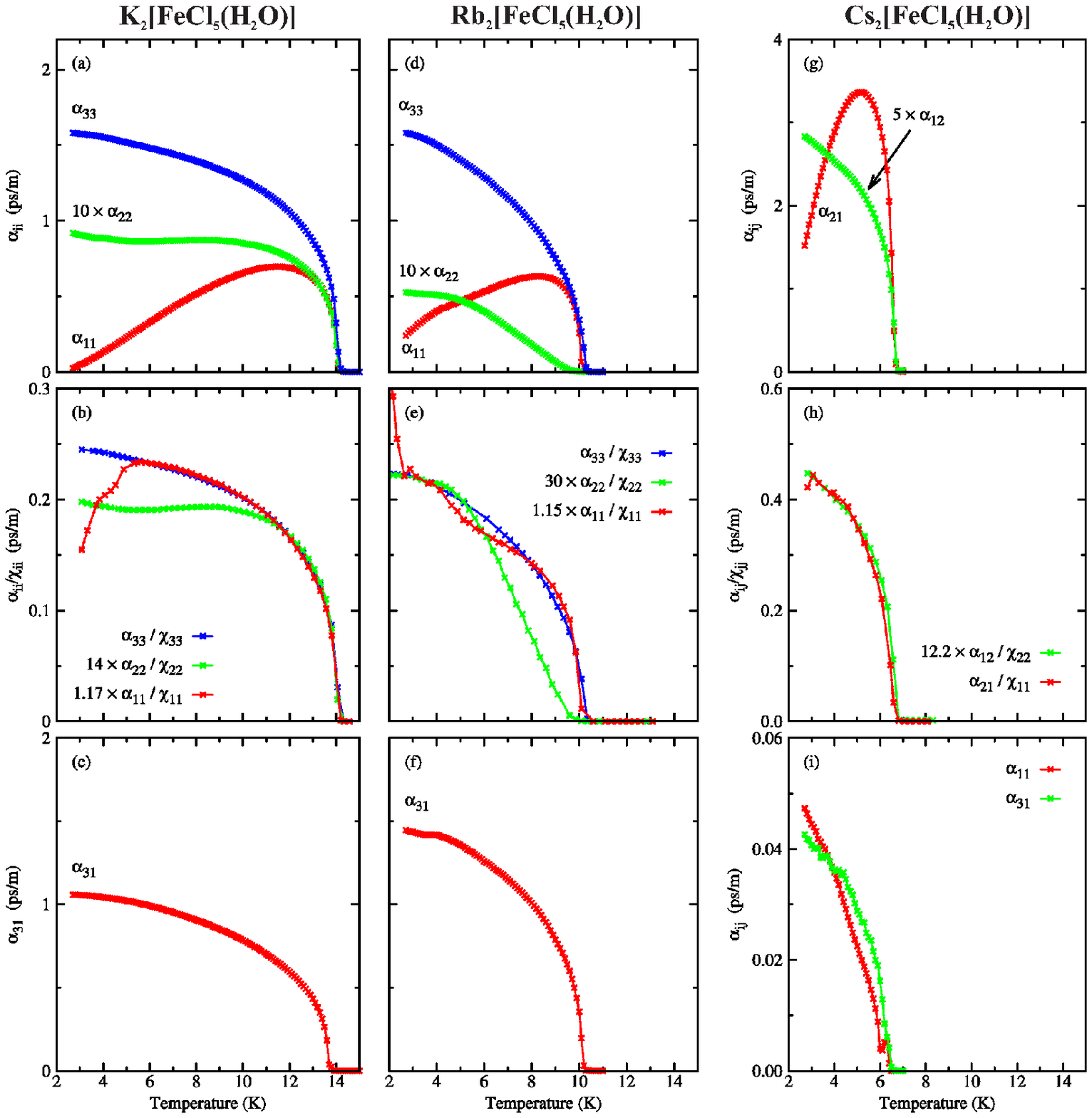}
\caption{Magnetoelectric properties of $A_2$[FeCl$_5$(H$_2$O)] with $A=$~K, Rb, Cs. (a), (d), (g) Temperature dependences of the linear magnetoelectric coefficients in the antiferromagnetic phases. (b), (e), (h) Temperature dependence of $\alpha_{||}/\chi_{||}$ for $\bi{B}$ parallel to the magnetic easy axis and of $\alpha_{\perp}/\chi_{\perp}$ for $\bi{B}$ perpendicular to the magnetic easy axis. (c), (f), (i) Temperature dependence of the linear magnetoelectric coefficients in the spin-flop phases.}
\label{figure9}
\end{figure}

The linear magnetoelectric tensors of K$_2$[FeCl$_5$(H$_2$O)] and Rb$_2$[FeCl$_5$(H$_2$O)] have a diagonal form with the non-zero components $\alpha_{11}$, $\alpha_{22}$ and $\alpha_{33}$ (related to unit vectors $\bi{e}_1, \bi{e}_2, \bi{e}_3$ parallel to the crystallographic axes $\bi{a}, \bi{b}, \bi{c}$, respectively). This result is compatible with the magnetic point-group symmetry $m'm'm'$, determined by neutron-diffraction measurements~\cite{Gabas1995a, Luzon2008a}. The temperature dependences and absolute values of corresponding tensor components $\alpha_{ii}(T)$ of both compounds are very similar, as can be seen in figure~\ref{figure9}\,(a),~(d). Below the N\'eel temperatures  $\alpha_{22}(T)$ and $\alpha_{33}(T)$ grow strongly and then saturate with decreasing temperature, while $\alpha_{11}(T)$ first reaches a broad maximum and finally approaches  zero for $T\rightarrow 0$. In contrast to the K-based and Rb-based compounds, Cs$_2$[FeCl$_5$(H$_2$O)] has only off-diagonal tensor components $\alpha_{21}$ and $\alpha_{12}$. The component $\alpha_{21}(T)$ has an analogous temperature dependence as $\alpha_{11}^{\rm K,Rb}(T)$ and the component $\alpha_{12}(T)$ behaves similar to $\alpha_{22}^{\rm K,Rb}(T)$ and $\alpha_{33}^{\rm K,Rb}(T)$, see figure~\ref{figure9}\,(g). The linear magnetoelectric tensors of $A_2$[FeCl$_5$(H$_2$O)] with $A=$~K, Rb, Cs calculated at representative temperatures are  displayed in table~\ref{tab:growth2}.

The spin-flop transitions for $\bi{B}||\bi{a}$ lead to a modification of the magnetoelectric responses in all $A_2$[FeCl$_5$(H$_2$O)] compounds. For $A=$~K or Rb, the component $\alpha_{11}$ is suppressed and $\alpha_{31}$ occurs. For $A=$~Cs, the component $\alpha_{21}$ is suppressed and $\alpha_{11}$ as well as $\alpha_{31}$ occur. Their temperature dependences are very similar to each other, see figures~\ref{figure9}\,(c), (f) and (i). The behaviour of the other tensor components in the spin-flop phases could not be determined, because in the used experimental setup it was not possible to apply a constant magnetic field $B>B_{\rm SF}$ parallel to $\bi{a}$, while applying a variable magnetic field parallel to $\bi{b}$ or $\bi{c}$.

For collinear antiferromagnetic magnetoelectrics, the temperature dependences of the linear magnetoelectric tensor components for $\bi{B}$ parallel ($\alpha_{||}$) and $\bi{B}$ perpendicular ($\alpha_{\perp}$) to the magnetic easy axis are characterized by the product of the corresponding magnetic susceptibilities $\chi_{||}(T)$ and $\chi_{\perp}(T)$ with the sublattice magnetization \mbox{$\bar{S}(T)$~\cite{Rado1,Rado2} via}
	\begin{equation}
	\label{eq:me4}
	\alpha_{||}(T) \propto \chi_{||}(T)\cdot\bar{S}(T), \quad \alpha_{\perp}(T) \propto \chi_{\perp}(T)\cdot\bar{S}(T).
	\end{equation}

As is shown in figure~\ref{figure9} the determined temperature-dependent $\alpha_{ij}(T)$ agree well to the expected behaviour according to equation~\ref{eq:me4}. For magnetic fields perpendicular to the magnetic easy axis ($\bi{B}||\bi{b}$ or $\bi{B}||\bi{c}$), the temperature dependences of the magnetoelectric responses are dominated by that of the sublattice magnetization $\bar{S}$, because the corresponding magnetic susceptibilities stay nearly constant below $T_{\rm N}$, see figure~\ref{figure2}. Therefore, $\alpha_{22}(T)$, $\alpha_{33}(T)$ for the K-based and Rb-based compound and $\alpha_{12}(T)$ for Cs$_2$[FeCl$_5$(H$_2$O)] have the form of an order parameter. For magnetic fields parallel to the magnetic easy axis ($\bi{B}||\bi{a}$), the temperature dependences of the magnetoelectric responses are dominated by the corresponding magnetic susceptibilities, which approach zero for $T\rightarrow 0$. Therefore, $\alpha_{11}(T)$ for the K-based and Rb-based compound and $\alpha_{21}(T)$ for Cs$_2$[FeCl$_5$(H$_2$O)] approach zero as well for $T\rightarrow 0$. The different temperature characteristics of the magnetoelectric responses are further illustrated in figures~\ref{figure9}\,(b),~(e) and (h), where the ratios $\alpha_{||}/\chi_{||}$ and $\alpha_{\perp}/\chi_{\perp}$ are displayed. For the calculation of the ratios $\alpha/\chi$, the magnetic volume susceptibilities in SI units $\chi^{\mathrm{SI}}=4\pi \chi^{\mathrm{CGS, mol}}/V_{\mathrm{mol}}$ are used. In most of the cases the expected temperature dependence of an order parameter is nicely reproduced. Partially, slight deviations occur in the low-temperature limit.  These can be explained by problems in determining the accurate absolute values of the electric polarization from the results of pyroelectric-current measurements, when the respective pyroelectric currents hardly exceed the background currents.

\begin{table}[t]
\centering
\begin{tabular}{cccc}
\hline
compound & $T$ (K) & $[\alpha_{ij}]_{\rm AF}$ (ps/m) & $[\alpha_{ij}]_{\rm SF}$ (ps/m) \\ 
\hline \\

K$_2$[FeCl$_5$(H$_2$O)] & \unit[11]{K} & $\left[\begin{array}{ccc}
 0.71 & \sim 0 & \sim 0\\
 \sim 0 & 0.08 & \sim 0\\
 \sim 0 & \sim 0 & 1.17
\end{array}\right]$ & $\left[\begin{array}{ccc}
 \sim 0 & \,\quad-\quad & -\\
 \sim 0 & - & -\\
  0.71 & -  & -
\end{array}\right]$ \\ \\

Rb$_2$[FeCl$_5$(H$_2$O)] & \unit[8]{K} & $\left[\begin{array}{ccc}
 0.63 & \sim 0 & \sim 0\\
 \sim 0 & 0.02 & \sim 0\\
 \sim 0 & \sim 0 & 0.98
\end{array}\right]$ & $ \left[\begin{array}{ccc}
 \sim 0 & \quad-\quad & -\\
 \sim 0 & - & -\\
  1.0 & -  & -
\end{array}\right] $ \\ \\

Cs$_2$[FeCl$_5$(H$_2$O)] & \unit[5]{K} & $ \left[\begin{array}{ccc}
 \sim 0 & 0.45 & \sim 0\\
 3.35 & \sim 0 & \sim 0\\
 \sim 0 & \sim 0 & \sim 0
\end{array}\right]$ & $\left[\begin{array}{ccc}
 0.03 & \quad-\quad & -\\
 \sim 0 & - & -\\
  0.02 & -  & -
\end{array}\right]$ \\ \\
\hline
\end{tabular}
\label{tab:growth2}
\caption{Linear magnetoelectric tensors of $A_2$[FeCl$_5$(H$_2$O)] with $A=$~K, Rb, Cs at representative temperatures. The tensors in the right column refer to the spin-flop phases for $B>B_{\rm SF}$ along $\bi{a}$, where only the components $\alpha_{i1}$ could be determined.}
\end{table}

As mentioned in the introduction, the magnetic structure of Cs$_2$[FeCl$_5$(H$_2$O)] is still unknown. With the knowledge of the anisotropy of the linear magnetoelectric effect of Cs$_2$[FeCl$_5$(H$_2$O)] it is, however, possible to deduce its magnetic space group by the following symmetry analysis.

\begin{figure}[t]
\includegraphics[width=\textwidth]{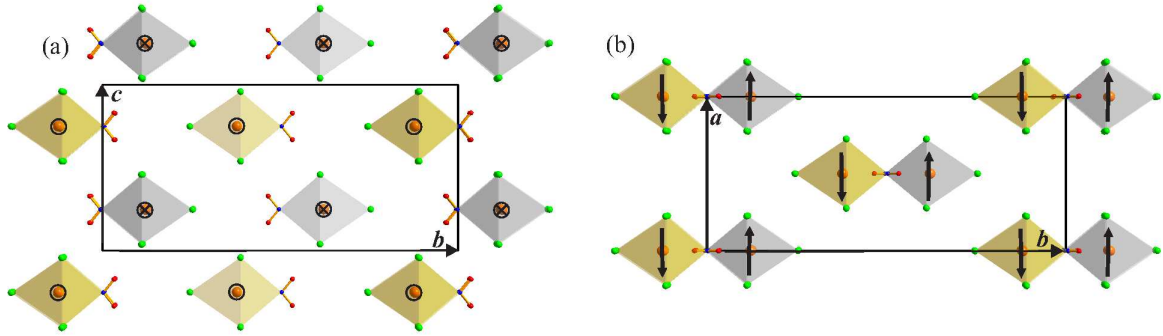}
\caption{Probable magnetic structure of Cs$_2$[FeCl$_5$(H$_2$O)] below $T_{\rm N}=\unit[6.8]{K}$, deduced from a symmetry analysis of its linear magnetoelectric tensor. The spin directions are along $\pm\bi{a}$ as indicated by $\circ$ versus $\times$ in the view along $\bi{a}$ (left) and by arrows in the view along $\bi{c}$. Along the zigzag chains running along $\bi{c}$ the spin direction alternates between $\pm\bi{a}$, whereas the spins of neighbouring chains are in phase.}
\label{figure10}
\end{figure}

Starting with the point group symmetry $\frac{2}{m}\frac{2}{m}\frac{2}{m}$ for the prototypic, paramagnetic phase of Cs$_2$[FeCl$_5$(H$_2$O)], the only possible magnetic point groups, compatible with the form of the linear magnetoelectric tensor, are $mm2$, $2'mm'$, $2'2'2$ and $mmm'$. The magnetic susceptibility measurements reveal that the spins are ordered antiferromagnetically below the N\'eel temperature with $\bi{a}$ as the magnetic easy axis. In addition, the results of the dielectric investigations show that in the antiferromagnetically ordered phase no electric polarization occurs in zero magnetic field. Therefore, the two polar groups $mm2$ and $2'mm'$ can be (almost certainly) excluded, as well as the ferromagnetic group $2'2'2$. Hence, the only remaining possibility is the magnetic point group $mmm'\mathrel{\hat=}\frac{2'}{m}\frac{2'}{m}\frac{2}{m'}$. Including the information about the space group $Cmcm\mathrel{\hat=}C\frac{2}{m}\frac{2}{c}\frac{2_1}{m}$ of the room-temperature crystal structure of Cs$_2$[FeCl$_5$(H$_2$O)], the (only possible) magnetic space group for its linear magnetoelectric phase is $C\frac{2'}{m}\frac{2'}{c}\frac{2_1}{m'}$, which corresponds to the magnetic space group number 63.5.515 according to~\cite{Litvin}. With this knowledge, a model for the magnetic structure of Cs$_2$[FeCl$_5$(H$_2$O)] can be derived. Starting with a spin at an arbitrary Fe site oriented along $\bi{a}$, the orientations of all the other spins within the structure follow by applying all the symmetry elements of the magnetic space group $C\frac{2'}{m}\frac{2'}{c}\frac{2_1}{m'}$. In figure~\ref{figure10} the resulting magnetic structure of Cs$_2$[FeCl$_5$(H$_2$O)] below $T_{\rm N}=\unit[6.8]{K}$ is displayed for one unit cell. The spin direction alternates between $\pm\bi{a}$ along the zigzag chains of the [FeCl$_5$(H$_2$O)]$^{2-}$ octahedra running along $\bi{c}$ and the spins of neighbouring chains are in phase. Therefore, the magnetic structure of the Cs-based compound is of the same type as that of the K-based and Rb-based compounds. The direction of the AFM zigzag chains is equivalent in the $Pnma$-type and  $Cmcm$-type structure, but the orientations of the magnetic easy axes differ by 90$^{\circ}$.

\section{Phase diagrams and conclusion}

\begin{figure}[t]
\includegraphics[width=\textwidth]{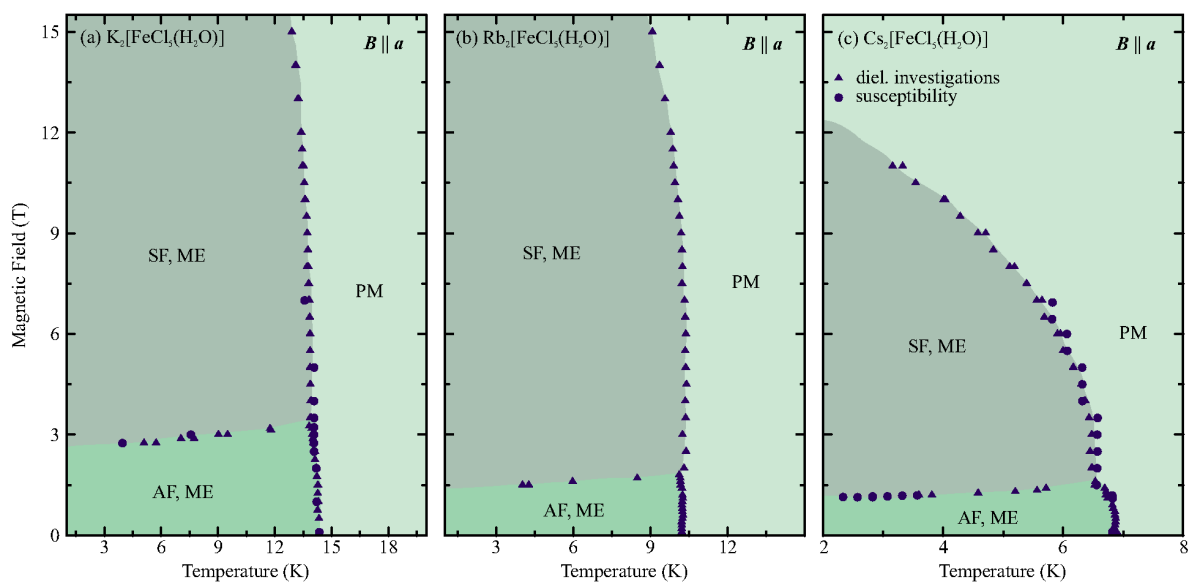}
\caption{Magnetic field versus temperature phase diagrams of K$_2$[FeCl$_5$(H$_2$O)], Rb$_2$[FeCl$_5$(H$_2$O)] and Cs$_2$[FeCl$_5$(H$_2$O)] for $\bi{B}$ parallel $\bi{a}$. The vertical phase boundaries between the paramagnetic (PM) and antiferromagnetic phases (AF) for $\bi{B}$ parallel $\bi{b}$ and $\bi{c}$ coincide within the experimental uncertainty with the respective ones for $\bi{B}$ parallel $\bi{a}$. The phase boundaries are based on the dielectric and magnetic measurements.}
\label{figure11}
\end{figure}

The present investigations reveal that all compounds $A_2$[FeCl$_5$(H$_2$O)] with $A=$~K, Rb, Cs are linear magnetoelectrics. By means of their magnetic-field versus temperature phase diagrams for $\bi{B}||\bi{a}$, their properties are summarized and compared in the following, see figure~\ref{figure11}. The phase boundaries are based on the results of the dielectric investigations and the magnetic-susceptibility measurements. The three phase diagrams are very similar. For each compound, there exists a paramagnetic phase (PM), a linear magnetoelectric, antiferromagnetically ordered phase (AF, ME) and a linear magnetoelectric spin-flop phase (SF, ME). With increasing ionic radius of the alkali metal in the crystal structure ($r_{\mathrm{Cs}}>r_{\mathrm{Rb}}>r_{\mathrm{K}}$), the zero-field transition temperature decreases from $T_{\rm N}^{\mathrm{K}}=\unit[14.3]{K}$ and $T_{\rm N}^{\mathrm{Rb}}=\unit[10.2]{K}$ to $T_{\rm N}^{\mathrm{Cs}}=\unit[6.8]{K}$. In the same way the critical spin-flop fields reduce from $B_{\mathrm{SF}}^{\mathrm{K}}=\unit[3.5]{T}$ and $B_{\mathrm{SF}}^{\mathrm{Rb}}=\unit[1.5]{T}$ to $B_{\mathrm{SF}}^{\mathrm{Cs}}=\unit[1.2]{T}$. The phase boundaries of the K-based and the Rb-based compound between the paramagnetic and the magnetically ordered phases hardly change with field, while that of  the Cs-based compound is strongly bent towards lower temperatures for increasing magnetic field strength. Figure~\ref{figure11} only displays the phase diagrams for $\bi{B}$ parallel to $\bi{a}$ because, apart from the absence of spin-flop transitions, the respective phase boundaries for $\bi{B}$ parallel to $\bi{b}$ and $\bi{c}$ essentially coincide with the phase boundaries for $\bi{B}$ parallel to $\bi{a}$ for all three compounds.

The magnetic point group $m'm'm'$, determined for K$_2$[FeCl$_5$(D$_2$O)] and Rb$_2$[FeCl$_5$(D$_2$O)] via neutron scattering~\cite{Gabas1995a, Luzon2008a} is consistent with the results of the present magnetoelectric investigations. For both compounds the linear magnetoelectric tensor has a diagonal form with the components $\alpha_{11}$, $\alpha_{22}$ and $\alpha_{33}$. This tensor form would also allow the point-group symmetry $222$ and $m'm'2$. Because $m'm'2$ is polar and no electric polarization is observed in zero magnetic field for both compounds, it can be excluded. The point group $222$, however, is like $m'm'm'$ consistent with the results of the present study. For Cs$_2$[FeCl$_5$(H$_2$O)] the linear magnetoelectric tensor has only the off-diagonal elements $\alpha_{12}$ and $\alpha_{21}$, which leads after a symmetry analysis to the magnetic point group $mmm'$ and magnetic space group $C\frac{2'}{m}\frac{2'}{c}\frac{2_1}{m'}$ for the magnetoelectric phase of this compound. Based on this result, the model shown in figure~\ref{figure10} is proposed for the magnetic structure of Cs$_2$[FeCl$_5$(H$_2$O)], which is of the same type as that of the K-based and Rb-based compound. However, the orientations of the magnetic easy axes in both structure types are different.

As mentioned in the introduction another member of the erythrosiderite-type family, (NH$_4$)$_2$[FeCl$_5$(H$_2$O)], has been established as new multiferroic material with a strong magnetoelectric coupling recently~\cite{Ackermann2013}. Although the alkali-based compounds investigated in the present work are structurally closely related to (NH$_4$)$_2$[FeCl$_5$(H$_2$O)], their magnetic properties are less complex and consequently they are only linear magnetoelectrics. The spin structure of (NH$_4$)$_2$[FeCl$_5$(H$_2$O)] exhibits an $XY$ anisotropy with a magnetic easy plane. Interestingly, this easy plane corresponds to the plane that is spanned by the two different orientations of the magnetic easy axes found in the K-based and Rb-based compound on the one hand and in the Cs-based compound on the other hand. Therefore, already slight crystal-chemical modifications within the erythrosiderite-type structure can change the orientation of the easy axes either by 90$^{\circ}$ or even transform the easy-axis anisotropy into an easy plane anisotropy. The fact that the easy-plane anisotropy allows the occurrence of a spiral spin structure suggests that such spin spirals may induce the multiferroicity in (NH$_4$)$_2$[FeCl$_5$(H$_2$O)] via the so-called inverse Dzyaloshinskii-Moriya mechanism~\cite{Sergienko2006}.

\section*{Acknowledgements}
We thank S.~Heijligen for measurements of the magnetic susceptibility.
This work was supported by the Deutsche Forschungsgemeinschaft via SFB 608 and through the Institutional Strategy of the University of Cologne within the German Excellence Initiative.

\end{document}